%% file: 1_maintext.tex
\documentclass[%
 reprint,
 amsmath,amssymb,
 aps,
floatfix,
]{revtex4-2}

\usepackage{graphicx}
\usepackage{dcolumn}
\usepackage{bm}
\usepackage{upgreek}
\usepackage{float}

\begin{document}

\preprint{APS/123-QED}

\title{Observation and control of hybrid spin-wave–Meissner-current transport modes} 
\author{M. Borst} 
\author{P.H. Vree} 
\author{A. Lowther} 
\author{A. Teepe} 
\author{S. Kurdi} 
\author{I. Bertelli} 
\author{B.G. Simon} 
\author{Y.M. Blanter} 
\author{T. van der Sar} 
\affiliation{Kavli Institute of Nanoscience, Delft University of Technology, 2628 CJ Delft, The Netherlands}

\
\date{\today}

\begin{abstract}
\input{sw_sc_abstract.tex}
\end{abstract}

\maketitle

\section{\label{sec:introduction}Introduction}
The ability to control the transport of spins and charges using metal electrodes is fundamental to information processing devices and an indispensable tool in quantum and condensed matter physics. While devices such as spin valves and transistors are based on transport of uncorrelated particles~\cite{Wolf2001}, the excitations of magnetic materials known as spin waves are emerging as promising alternative information carriers~\cite{Chumak2015a}. These collective spin excitations provide new opportunities for realizing analog or binary device functionality based on their wave nature, non-reciprocal transport properties, and low intrinsic damping~\cite{Barman2021}. 

Control of spin-wave transport is possible by heavy-metal electrodes that enable modulation via the spin-Hall effect~\cite{Sinova2015,Cornelissen2015b,Hamadeh2014} or by auxiliary magnetic materials that modify the spin-wave spectrum~\cite{Yu2019,Chen2018}. However, metallic gates can also introduce additional spin-wave damping because of uncontrolled spin pumping or spin-wave-induced eddy currents~\cite{Hamadeh2014, Sun2013, Bertelli2021}. Furthermore, the diamagnetic response of normal metals is dominated by Ohmic resistance, precluding effective stray-field control of the spin-wave spectrum. 

An attractive approach for strong, low-damping spin-wave modulation is to use superconducting electrodes. Superconductors are materials with zero electrical resistivity and a strong diamagnetic response that enables creating magnetic shields, magnetic lenses, and circuits such as quantum bits and quantum interference devices~\cite{Zhang2012, Blais2021}. Spin-wave spectroscopy measurements have demonstrated that superconducting strips on magnetic films can alter the spin-wave spectrum through the backaction of induced currents~\cite{Golovchanskiy2020} or the interaction with Abrikosov vortices~\cite{Dobrovolskiy2019}. Recently, it was proposed to harness the diamagnetism of a superconductor to create the spin-wave equivalents of optical mirrors and cavities~\cite{Yu2022}. Being able to image and control spin waves as they travel underneath superconducting electrodes would enable insight into the nature of the spin wave–superconductor interaction and unlock opportunities to control the propagation, dispersion, and refraction of spin waves.

In this work, we develop, image, and study temperature-, field-, and laser-tunable spin-wave transport enabled by a superconducting strip on a thin-film magnetic insulator (Fig.~1A). We use magnetic resonance imaging based on nitrogen-vacancy (NV) spins in diamond~\cite{Rondin2014, Zhou2021, Bertelli2020} to study the spin waves as they travel underneath the optically opaque superconductor. We demonstrate strong, temperature- and field-controllable shifts of the spin-wave lengths caused by the diamagnetic response of the superconductor, providing a striking visualization of the temperature-induced change of the kinetic inductance and London penetration depth. We explain our observations by incorporating the backaction of the spin-wave-induced Meissner currents into the Landau-Lifshitz-Gilbert (LLG) equation, through which we find the dispersion describing these hybrid spin-wave–Meissner-current modes. The agreement of this model with our data enables quantitative measurements of the London penetration depth as a function of temperature. Furthermore, we demonstrate that creating a local hot spot in the superconductor using a focused laser induces spin-wave refraction at target sites. The ability to image and control the spin-wave dispersion and refraction using Meissner currents provides a new window into magnet-superconductor interaction and opens the way for circuit elements such as spin-wave mirrors, gratings, filters, crystals and cavities. 

\section{Imaging hybrid spin-wave – Meissner-current transport modes using spins in diamond}

\begin{figure*}
\includegraphics[scale=0.97]{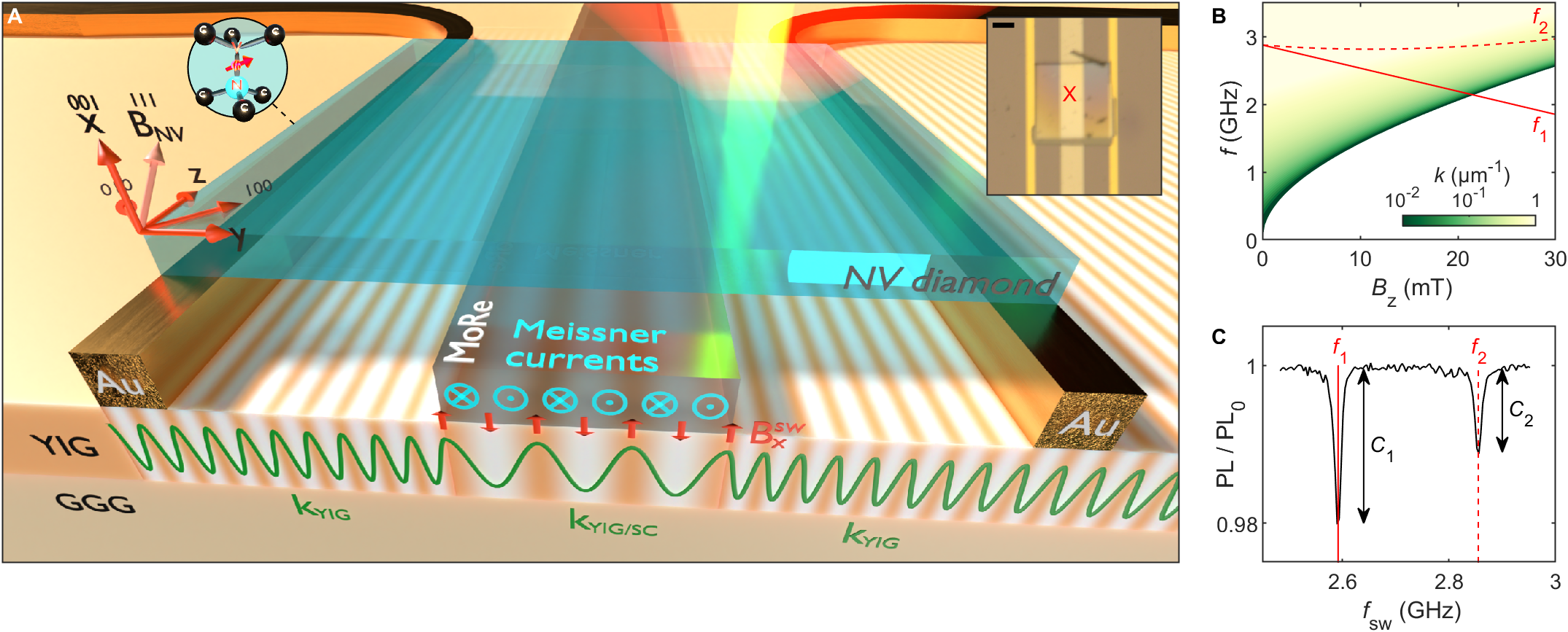}
\caption{\label{fig:blender}\textbf{Magnetic resonance imaging of hybridized spin-wave–Meissner-current transport modes.} (\textbf{A}) Overview of the experiment. A gold (Au) microstrip excites spin waves with wavevector $\mathbf{k}=k\hat{\mathbf{y}}$ in a 245-nm-thick film of yttrium iron garnet (YIG). The spin waves travel towards a molybdenum rhenium (MoRe) superconducting strip (width $W=30\ \upmu\text{m}$), thickness $t=140\text{ nm}$) where their stray fields induce Meissner currents that act back on the spin waves, shifting their wavenumber to $k_\text{YIG/SC} < k_\text{YIG}$. We image the waves underneath and next to the superconductor by their microwave magnetic stray fields using a $\sim$ 20-nm-thick layer of nitrogen-vacancy (NV) spins implanted at $\sim 0.07\ \upmu$m below the bottom surface of a $100\times 100 \times 5\ \upmu\text{m}^3$ diamond membrane placed on top of the sample. A magnetic field $\mathbf{B}_\text{NV} = B_x\hat{\mathbf{x}} + B_z \hat{\mathbf{z}}$ applied at $\theta = 54$ deg. with respect to the x-axis yields an in-plane YIG magnetization along $\hat{\mathbf{z}}$ for the small fields applied and directional Damon-Eshbach spin-wave excitation. Inset: optical micrograph of NV-diamond and Au and MoRe strips on the YIG. Scale bar: $30\ \upmu\text{m}$. Au thickness: $200\text{ nm}$. (\textbf{B}) YIG dispersion (color map) and NV electron spin resonance (ESR) frequencies (red lines) as a function of the in-plane field component $B_z = B_\text{NV} \cos(\theta)$. $f_{1(2)}$ denotes the ESR frequency of NV spins with zero-field quantization axis aligned (misaligned) with $B_\text{NV}$. The intersection of the ESR frequencies with the spin-wave dispersion sets the detectable spin-wavenumbers $k$. (\textbf{C}) Optically detected NV ESR spectrum at $B_z=10\text{ mT}$, at a location denoted by the red cross in the inset of (A). The ESR contrast $C_{1(2)}$ results from interference between the microstrip field and spin-wave field, enabling spatial mapping of the spin-wave fronts.}
\end{figure*}

Our system consists of a thin film of yttrium iron garnet (YIG) — a magnetic insulator with low spin-wave damping~\cite{Chumak2015a} — equipped with gold microstrips for spin-wave excitation and a molybdenum-rhenium superconducting strip for spin-wave modulation (Fig.~1A). To image the spin waves, we place a diamond membrane that contains a thin layer of nitrogen vacancy (NV) sensor spins on top of the sample (Fig.~1A, Materials and Methods, Fig.~S1)~\cite{Bertelli2020}. These spins detect the spin waves by their microwave magnetic stray fields, enabling imaging through optically opaque materials~\cite{Bertelli2021}. The sample is embedded in a variable temperature cryostat with a base temperature of 5.5 K and free-space optical access to read out the NV sensor spins. 

NV centers are atomic defects in the diamond carbon lattice with an $S=1$ electron spin~\cite{Rondin2014}. The sensitivity of the NV spin to magnetic fields, combined with its optical spin readout and excellent spin coherence, has enabled widespread sensing applications in fields ranging from condensed-matter science to geology and biophysics~\cite{Casola2018,Schirhagl2014, Glenn2017}. Here we use the sensitivity of the NV spins to microwave magnetic fields to image the spin waves in the YIG film~\cite{Zhou2021, Bertelli2020, Andrich2017, Wolfe2014a}. When resonant with an NV electron spin resonance (ESR) frequency, the stray field of the spin waves drives transitions between the NV spin states that we detect through the spin-dependent NV photoluminescence under green-laser excitation.

\begin{figure*}[!ht]
\includegraphics[scale=1.15]{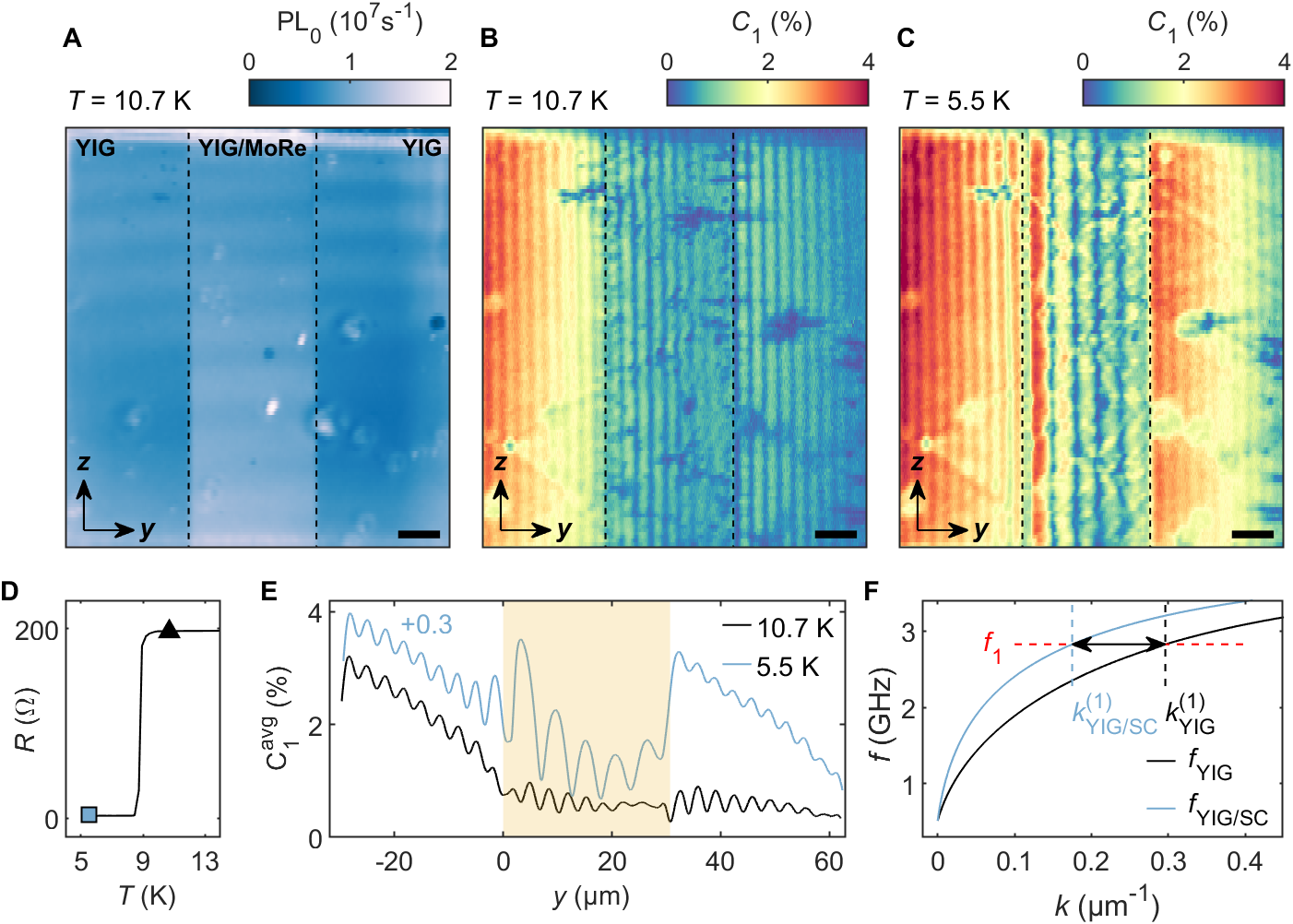}
\caption{\label{fig:spatial_map} \textbf{Magnetic resonance imaging of spin waves above and below the superconducting transition temperature.} (\textbf{A}) Spatial map of the NV photoluminescence $\text{PL}_0$ in the absence of microwaves, showing the MoRe strip (between the vertical dashes). Scale bar: $10\ \upmu\text{m}$. (\textbf{B}, \textbf{C}) Spatial maps of the NV electron spin resonance contrast $C_1$ above (B) and below (C) the superconducting transition temperature of $T_c=8.7\text{ K}$, at $T=10.7\text{ K}$ and $T=5.5 \text{ K}$ respectively. The Au microstrip exciting spin waves is located just outside the left edge of the imaged area. Above (below) $T_c$, the wavelength is unaffected (lengthened) by the MoRe strip. (\textbf{D}) DC resistance R of the MoRe strip as a function of temperature $T$, with markers indicating the resistance of the film during the measurements of (B), triangle, and (C), square. (\textbf{E}) Data from (B) and (C) averaged over the z-direction, with the MoRe strip indicated by yellow shading. \textbf{(F)} Calculated spin-wave dispersion $f_\text{YIG}(k)$ for bare YIG and $f_\text{YIG/SC}(k,\lambda_L)$ for YIG covered by a superconducting film with London penetration depth $\lambda_L=400\text{ nm}$. The superconductor shifts the dispersion upwards by $f_\text{SC}(k,\lambda_L)$, which manifests as a reduction in the wavenumber at the NV frequency $f_1$ from $k_\text{YIG}^{(1)}$ to $k_\text{YIG/SC}^{(1)}$ as indicated by the dashed lines.}
\end{figure*}

We apply a magnetic bias field to tune the NV ESR frequency into resonance with spin waves of different spin-wave lengths (1B). By orienting the field along one of the four possible crystallographic NV orientations (Fig.~1A), we split the ESR frequency of this ‘field-aligned’ NV ensemble ($f_1$) off from that of the three other NV ensembles ($f_2$) as shown in the optically detected resonance spectrum of Fig.~1C. Alternatively, we apply the field in-plane along $\hat{\mathbf{z}}$ to enable measurements at different frequencies at a given magnetic field. Because the applied magnetic fields are much smaller than the YIG saturation magnetization, the YIG magnetization lies predominantly in-plane along $\hat{\mathbf{z}}$ for both field orientations (Damon-Eshbach geometry). 

\begin{figure*}[!ht]
\includegraphics[scale=1.15]{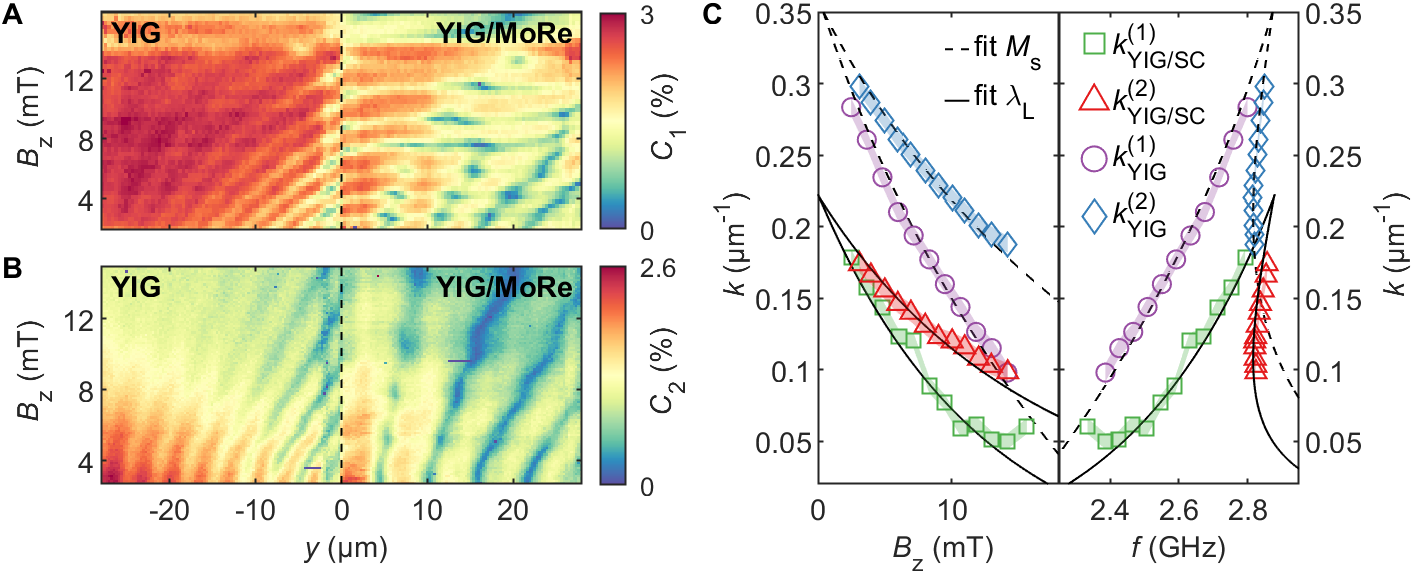}
\caption{\label{fig:field_dep} \textbf{Magnetic field dependence of the spin-wave dispersion in the magnet-superconductor hybrid}. (\textbf{A}, \textbf{B}) Spatial line traces of the NV ESR contrast $C_{1(2)}$, where the subscript identifies the field-aligned (misaligned) NV ensemble, as a function of magnetic field $B_z$. Spin waves excited in the bare-YIG region $(y<0)$ by the left Au microstrip (outside the imaged area) travel towards and then underneath the superconducting strip $(y>0)$, changing their wavelength. Interference with secondary spin waves excited at the MoRe strip edge (at $y=0$) due to inductive coupling between Au and MoRe strips yields a beating pattern along $B_z$ for $y>0$. $T=5.5\text{ K}$. The drive frequency is adjusted at each $B_z$ to maintain resonance with the NV ESR transitions. (\textbf{C}) Spin-wave number as a function of field (left panel) and frequency (right panel), extracted from the data in (A) and (B) via Fourier transformation (Fig.~S2). The $k^{(i)}$  are the wavenumbers measured with the field-aligned ($i=1$) and misaligned ($i=2$) NV ensembles. The error bars (indicated by shading) are determined by the inverse of the spatial sampling range in the $y$-direction. We determine the saturation magnetization $M_s$ by fitting the data in the bare-YIG region, and the London penetration depth $\lambda_L$ by fitting the data in the YIG/MoRe region using our YIG/SC model (Supplementary Text, Section 2).}
\end{figure*}

To demonstrate the spin-wave modulation capabilities of our superconductor, we image the spin-wave transport above and below the MoRe superconducting transition temperature $T_c=8.7 \text{ K}$(Fig.~2A-D). We generate NV-resonant spin waves with wavevector $\mathbf{k}=k\hat{\mathbf{y}}$ by applying a microwave current at NV frequency $f_1$ to the gold microstrip that is located just left outside the imaging area. The interference between the microwave magnetic stray field generated by these spin waves and the direct microstrip field leads to a spatial standing-wave modulation of the NV ESR contrast~\cite{Zhou2021,Bertelli2020}. Crucial for our measurements, this interference effect enables a straightforward extraction of the spin-wave length. The spatial map of the ESR contrast $C_1$ at $T=10.7\text{ K}$ (above $T_c$) shows spin waves traveling towards and then underneath the MoRe strip without a change in wavelength (Fig.~2B). In contrast, the spin-wave length increases almost twofold when the strip is cooled into its superconducting state at $T=5.5\text{ K}$ (Fig.~2C). Averaging the maps along $\hat{\mathbf{z}}$ (Fig.~2E) highlights the spatial homogeneity of the wavelength change.

We explain the superconductor-induced change of the spin-wave length by developing an analytical expression for the spin-wave dispersion in a magnet-superconductor thin-film hybrid. In this model, building on the formalism developed in~\cite{Yu2022}, the spin waves induce AC Meissner currents that are governed by the London penetration depth $\lambda_\text{L}$ of the superconductor. These currents, in turn, generate a magnetic field that acts back on the spin waves. By integrating this field self-consistently into the Landau-Lifshitz-Gilbert (LLG) equation, we find that the spin-wave dispersion shifts upwards in frequency as
\begin{equation}
    f_\text{YIG/SC}(k,\lambda_\text{L}) = f_\text{YIG}(k)+f_\text{SC}(k,\lambda_\text{L}),
\end{equation}
where $f_\text{YIG}(k)$ is the bare-YIG spin-wave dispersion (SI) and
\begin{equation}
    f_\text{SC}(k,\lambda_L) \approx \gamma \mu_0 M_s k t r \dfrac{1-e^{-2h/\lambda_L}}{(k\lambda_L+1)^2 - (k\lambda_L-1)^2 e^{-2h/\lambda_L}} 
\end{equation}
is the superconductor-induced shift (Supplementary Text, Section S1). Here, $M_s$ is the YIG saturation magnetization, $t$ = 245 nm is the YIG thickness, $h$ = 140 nm is the superconductor thickness, $\gamma$ = 28 GHz/T is the electron gyromagnetic ratio, $\mu_0$ is the vacuum permeability, and $r$ is a dimensionless factor associated with the YIG thickness and spin-wave ellipticity. The approximation holds when the kinetic inductance dominates the impedance, as is the case for our superconducting strip (Supplementary Text, Section S1), and when $k^2 \lambda_L^2 \ll 1$. A more general expression is given in the Supplementary Text. The dispersion shift $f_\text{SC}(k,\lambda_L)$ is maximal when $\lambda_L \rightarrow 0$, in which case the superconductor perfectly screens the spin-wave stray field. The calculated bare-YIG and hybridized YIG/MoRe spin-wave dispersions are compared in Fig.~2E. The upwards frequency shift underneath the superconductor manifests as a reduction in wavenumber of the NV-resonant spin waves detected in our experiments (Fig.~2E).

\begin{figure*}[!ht]
\includegraphics[scale=1.13]{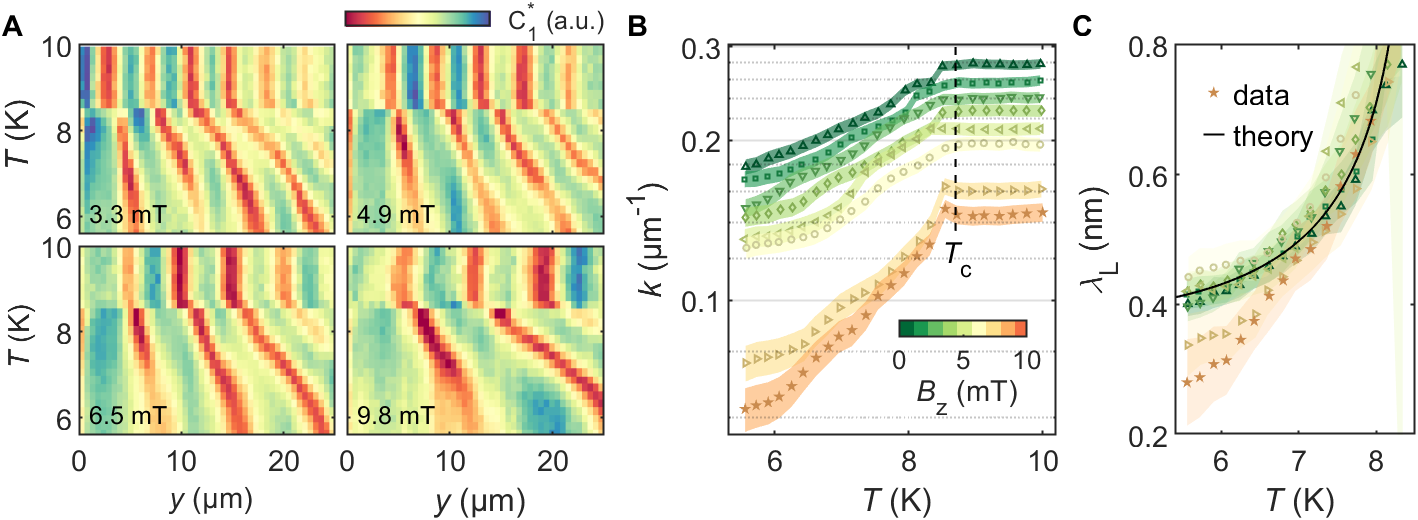}
\caption{\label{fig:temp_dep} \textbf{Temperature tunability of the hybrid spin-wave–Meissner-current dispersion and extraction of the London penetration depth.} (\textbf{A}) Spatial line traces of the NV ESR contrast $C_1$ across the YIG/MoRe region as a function of temperature, showing the continuous change of the spin wavelength underneath the MoRe strip, for different in-plane magnetic fields $B_z$. The data are linearly detrended along $y$. Above the superconducting phase transition there is no temperature dependence of the wavelength, indicating the absence of the Meissner effect. (\textbf{B}) Spin-wave numbers $k$ extracted from data in (A) and from additional data in Supplementary Fig.~S4 as a function of temperature. The colors indicate the different magnetic field values $B_z$. (\textbf{C}) London penetration depth $\lambda_L$ of the MoRe film as a function of temperature, extracted from the data in (B) via our YIG/SC model. Black line: fit of the temperature dependence of $\lambda_L (T)$ from which we extract $T_c=8.7\text{ K}$ and $\lambda_L^0=380\text{ nm}$. The colors indicate the different values of $B_z$ as in (B).}
\end{figure*}

\section{Temperature- and field-dependence of the spin-wave dispersion and extraction of the London penetration depth.} 

We characterize the magnetic-field dependence of the spin-wave dispersion underneath the superconductor and use it to extract the London penetration depth $\lambda_L$ at the $T=5.5 \text{ K}$ base temperature of our cryostat. Spatial line traces of the NV ESR contrast across the strip show the dependence of the spin-wave length on the applied magnetic field for the field-aligned (Fig.~3A) and misaligned (Fig.~3B) NV ensembles. In both measurements, we adjust the drive frequency at each magnetic field to maintain resonance with the NV ESR frequency. We extract the spin-wave numbers in the bare-YIG and YIG/MoRe regions separately by Fourier transformation (Fig.~S2), and plot these as a function of field and frequency in Fig.~3C and Fig.~3D. A similar measurement with the bias field applied in-plane along $\hat{\mathbf{z}}$ shows that Meissner screening of the bias field does not play a significant role in the wavelength shift (Fig.~S3). 

From the field-dependence of the extracted spin-wave numbers in the bare-YIG region, we extract the YIG saturation magnetization $M_s=194(1)\text{kA/m}$ (Supplementary Text, Section S2) in agreement with previous low-temperature measurements~\cite{Knauer2023}. We then use $M_s$ as a fixed parameter to fit the field-dependence of the spin-wave numbers underneath the superconducting strip using the hybridized YIG/MoRe spin-wave dispersion (Eq. 1). From this fit we extract a London penetration depth $\lambda_L=405(10)\text{ nm}$ at $T=5.5\text{ K}$, which agrees well with static-field nano-squid measurements~\cite{Shishkin2020}. 

The temperature dependence of the London penetration depth provides a powerful tool for tuning the spin-wave length. To demonstrate this, we image the spin waves in the YIG/MoRe region while sweeping through $T_c$ at different magnetic fields (Fig.~4A, Fig.~S4). The extracted spin-wave number $k$ is shown in Fig.~4B, with the color indicating the in-plane component of the magnetic field. We observe that $k$ changes continuously with temperature over the superconducting phase transition in the YIG/MoRe region while remaining unchanged in the bare-YIG region (Fig.~S5). We note that we do not observe global heating of the superconductor due to our excitation laser (Fig.~S6). Using our model, we extract the London penetration depth $\lambda_L(T)$ for every observed value of $k$ (Fig.~4C, Supplementary Text, Section S2). We find that almost all data collapse onto a single curve described by $\lambda_L = \lambda_L^0\left[(1-T/T_c)^4\right]^{-1/2}$~\cite{Tinkham2004}, with $T_c=8.7\text{ K}$ and $\lambda_L^0=380\text{ nm}$. The exceptions occur when the spin-wave length $\lambda_\text{sw}=2\pi/k$ becomes comparable to the width of the MoRe strip. Here, our approximation of the superconducting strip by an infinite film breaks down. These results highlight that imaging the hybridized spin-wave–Meissner-current transport modes is a powerful tool for extracting the temperature dependence of the London penetration depth. 

\section{Local control of spin-wave transport by laser-induced spin-wave refraction}
\begin{figure}
\includegraphics[scale=1.1]{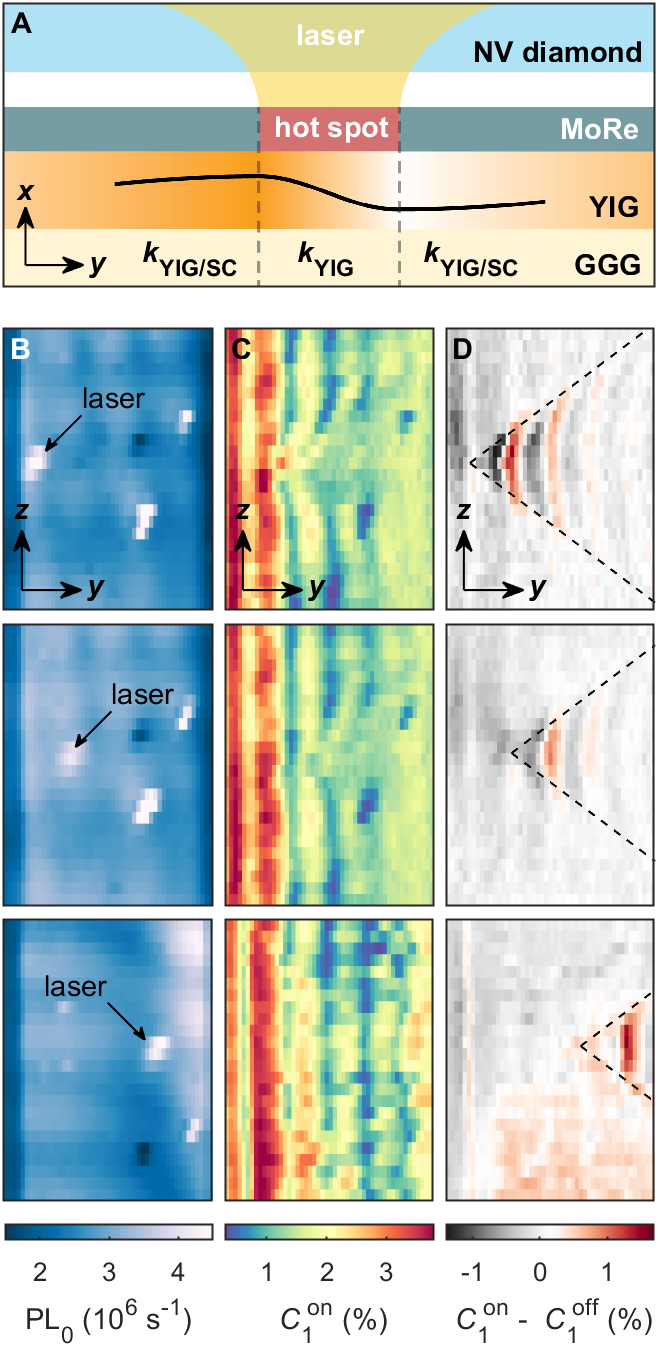}
\caption{\label{fig:scattering} \textbf{Laser-induced spin-wave refraction at target locations.} (\textbf{A}) Schematic illustration of a laser-induced scattering spot. By shining an auxiliary 594 nm laser on the sample, we create a hot spot in the MoRe strip that locally alters the effective refractive index governing the spin-wave propagation. (\textbf{B}) Scanning confocal microscope images of the NV photoluminescence at $T=5.5\text{ K}$, with the auxiliary laser focused onto the MoRe strip at three different locations indicated by the arrows. Scale bar: $10\ \upmu\text{m}$. (\textbf{C}) Spatial maps of the NV ESR contrast with auxiliary laser turned on ($C_1^\text{on}$), showing spin waves in the YIG/MoRe region that scatter on the laser spot. (\textbf{D}) Background-subtracted ESR contrast, highlighting the laser-induced spin-wave scattering, obtained by subtracting the ESR contrast with the auxiliary laser turned off $C_1^\text{off}$ from the measurements in (C).}   
\end{figure}

Thus far, we have demonstrated dispersion engineering through global control of temperature and magnetic field. We now show that the creation of a hot spot in the superconductor using a focused laser enables local manipulation of the spin-wave transport by tuning the effective refractive index (Fig.~5A). To do so, we couple an auxiliary, orange laser into our setup and focus it at target sites on the superconductor (Fig.~5B). The laser spot is visible through the locally enhanced NV photoluminescence. Spatial measurements of the NV ESR contrast $C_1^\text{on}$ with the auxiliary laser on (Fig.~5C) show the spin-wave scattering patterns induced by the local hot spot. The reduction in amplitude behind the hot spot indicates destructive interference between the scattered and incident spin waves. Subtracting a reference measurement with the auxiliary laser turned off (Fig.~5D) highlights the angular profile of the scattered spin-wave patterns. 

The characteristic ‘caustic’ angles observed in these scattered patterns (dashed lines in Fig.~5D) result from the highly anisotropic dipolar spin-wave dispersion~\cite{Gieniusz2013}. Tracing the patterns to their origin shows that the scattering site is tightly confined to the laser location. Presumably, the laser locally breaks the superconductivity, inducing a local change in the magnetic environment seen by the spin waves, leading to local spin-wave refraction akin to defect-controlled spin-wave scattering~\cite{Gieniusz2013}. The ability to optically induce spin-wave refraction at target sites could be used to create devices such as gratings or magnonic crystals~\cite{Chumak2014} and enable spin-wave manipulation via optical switching of flux-focusing regions in the superconducting strip.

\section{Conclusions}
We demonstrated local measurements of hybridized spin-wave – Meissner-current transport modes in a magnetic thin film equipped with a superconducting gate. The wavelength is tunable by temperature and field, enabling efficient phase-shifting of the spin-wavefronts and a striking in-situ visualization and quantitative extraction of the London penetration depth as a function of temperature. Our measurements did not reveal effects caused by Abrikosov vortices, the distribution of which is presumably strongly influenced by our focused excitation laser as demonstrated by magneto-optical~\cite{Veshchunov2016} and wide-field NV-imaging experiments~\cite{Lillie2020}. The presented microwave magnetic imaging of the spin-wave transport modes in a YIG/MoRe heterostructure shows the versatility of superconducting gates for spin-wave manipulation, enables determining the temperature-dependent London penetration depth, and opens new opportunities for creating wave-based circuit elements such as filters, mirrors, and cavities.

\section*{Author contributions}
\textbf{MB} designed and fabricated the sample. \textbf{BS} and \textbf{MB} fabricated the diamond membranes. \textbf{AL} and \textbf{SK} developed the membrane placement technique. \textbf{MB} constructed the experimental setup with help from \textbf{AL}. \textbf{MB}, \textbf{PV} and \textbf{AL} performed the measurements with help from \textbf{IB}. \textbf{MB} and \textbf{PV} analyzed the data with help from \textbf{AT} and \textbf{SK} and feedback from \textbf{TS}. \textbf{TS} and \textbf{YB} developed the theory. \textbf{MB} wrote the manuscript with guidance and redaction of \textbf{TS}, and input from all co-authors. \textbf{TS} conceived the experiment and supervised the project.

\begin{acknowledgments}
We thank prof. A.F.~Otte for commenting on the manuscript and prof. T.~Yu and prof. G.E.W.~Bauer for valuable discussions. This work was supported by the Dutch Research Council (NWO) under awards VI.Vidi.193.077, NGF.1582.22.018, and OCENW.XL21.XL21.058.  
\end{acknowledgments}

\clearpage
\bibliography{export.bib}

\end{document}


\beginsupplement 

\begin{center}
\textbf{{\huge Supplementary material for}}\break\break

\textbf{{\Large Observation and control of hybrid spin-wave–Meissner-current transport modes }}\break\break

{ {\large M.~Borst}$^{1}$, {\large P.H.~Vree}$^{1}$, {\large A.~Teepe}$^{1}$, {\large S.~Kurdi}$^{1}$, {\large I.~Bertelli}$^{1}$, {\large B.G.~Simon}$^{1}$, {\large Y.M.~Blanter}$^{1}$, \large T.~van der Sar}$^{1}$\break

\small{
$^1$Department of Quantum Nanoscience, Kavli Institute of Nanoscience, Delft University of Technology, 2628 CJ Delft, The Netherlands}
\end{center}
\tableofcontents

\section{Materials and Methods}
\subsubsection{Sample fabrication}
We use commercially available (Matesy GmbH) 245 nm thick yttrium iron garnet (YIG) films, grown on a gadolinium gallium garnet (GGG) substrate. To create the metal strips, we spin a positive PMMA bilayer e-beam resist and transfer our patterns to the chip with e-beam lithography. For the superconducting strip, we sputter 140 nm of molybdenum rhenium (MoRe) in an Alliance Concept AC450. For the gold microstrips, we evaporate 10 nm Ti followed by 140 nm Au in an AJA e-beam evaporator. Liftoff was done using acetone.

\subsubsection{Diamond membrane fabrication and placement}
To fabricate the diamond platelet with a dense nitrogen-vacancy ensemble, we purchase 4 x 4 x 0.5 mm$^3$ electronic-grade diamond plates (Element 6 Inc.) and have them cut and polished into 2 x 2 x 0.05 mm$^3$ platelets (Almax Easylabs). Nitrogen ions are implanted at an energy of 54 keV and density of 105 µm-2 (Innovion), after which the diamond is annealed in vacuum, resulting in an estimated nitrogen vacancy center density of 10$^3$ µm$^{-2}$  at a depth of approximately 70 nm below the surface \cite{Pezzagna2010}. 

Next, we reshape this platelet to host many 100 x 100 x 5 µm$^3$ diamond membranes. To do so, we first cover the diamond with a 50 nm Ti layer by e-beam evaporation, into which the membrane pattern is etched by a SF6/He plasma following e-beam lithography. The Ti layer serves as a hard mask for the next step, in which the membrane pattern is reactive-ion etched 5 µm into the diamond by an O2 plasma. We turn the diamond over and etch the back almost all the way through with the same O2 plasma, until the diamond membranes only remain attached to the host by a small holding bar. Finally, we remove the Ti with a hydrofluoric acid dip. 

To place these diamond membranes on a sample of interest, we glue the side of the carrier diamond to an elastic metal holding bar and mount it to an XYZ-manipulator, so that it can be positioned above the sample with ~µm accuracy. Next, we use a thin metal needle, attached to another XYZ-manipulator, to push the membrane on the sample by breaking its holding bar. This process is displayed in Fig.~S1 by a series of images. The membrane sticks to our sample as-is and did not displace under typical mechanical disturbances to the sample, such as wire bonding, vertical mounting in our cryostat, and slip-stick positioning of the sample.

\begin{figure}[!ht]
\includegraphics{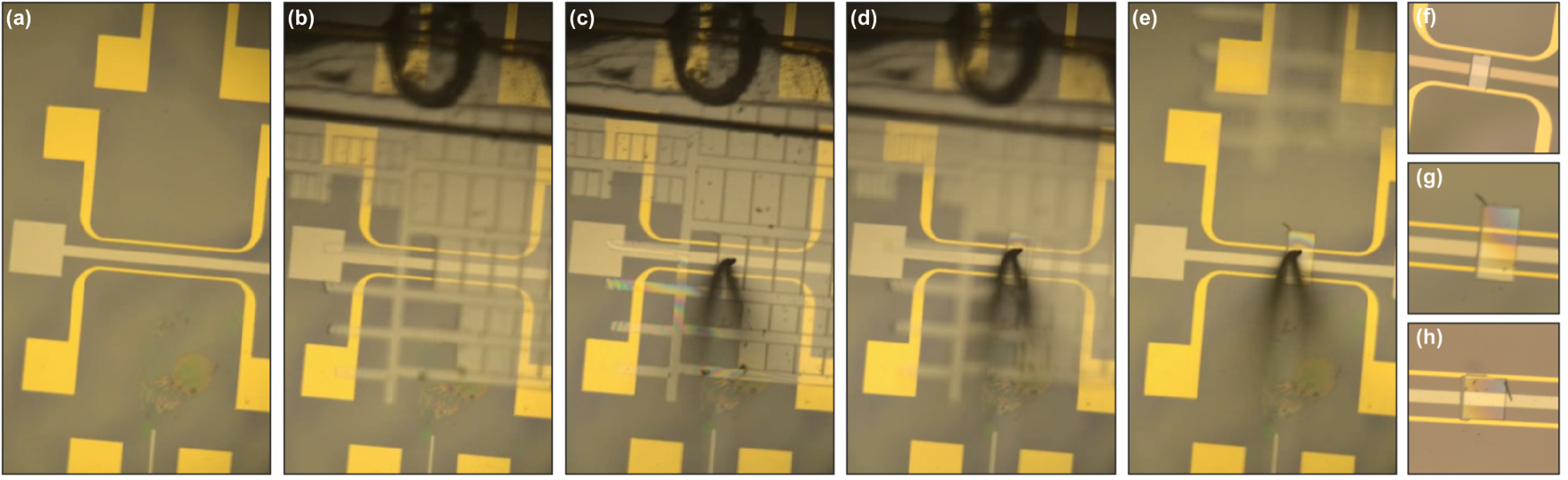}
\caption{\textbf{Diamond membrane tipping process for NV magnetometry.} (a) The sample region of interest is selected and brought into optical focus. (b) A carrier diamond hosting many NV sensing membranes is positioned above the sample area. The carrier diamond is glued to an elastic metallic spring, which is attached to an XYZ-micromanipulator for positioning. (c) A thin metallic needle, attached to a second XYZ-micromanipulator, is positioned above the selected membrane. (d) The membrane is tipped out with the needle. (e) After tipping, the membrane is brought into its final position by careful manipulation with the needle. (f-h) Various tipped-out membranes on top of devices of interest. }
\label{fig:S1}
\end{figure}

\subsubsection{Photoluminescence measurements}
We perform nitrogen vacancy center photoluminescence measurements in a closed cycle cryostat (Montana Cryostation S100) with optical access, through a room-temperature-stabilized NA = 0.85 microscope objective. We position the sample globally using slip-stick positioners (Attocube ANPx101, ANPz101), and use a fast-steering mirror (Newport FSM300) to scan a 520 nm continuous-wave laser (Coherent Obis 520LX) over our sample. We detect the corresponding NV photoluminescence by an avalanche photodiode (Excelitas SPCM-AQRH-13), after 600 nm longpass filtering the optical signal. The magnetic field used to magnetize our sample and set the NV resonance frequencies is applied by a large permanent magnet outside the sample chamber, in a home-built positioning system that has rotational (Zaber T-RS60) and translational (Zaber X-LRT0250AL-E08C) degrees of freedom. A microwave generator (Windfreak SynthHDv2) is used to apply microwave signals to the excitation stripline to excite spin waves and drive the NV center ensemble spins.

\section{Spin-wave dispersion underneath a superconducting thin film.}

Here we derive the spin-wave dispersion for the magnet-superconductor hybrid (Eq. 1 of the main text). We do so by integrating the magnetic field generated by the spin-wave-induced Meissner currents self-consistently into the linearized Landau-Lifshitz-Gilbert (LLG) equation.

We consider a superconducting film located between $0<x<h$ on top of a magnetic film located between $-t<x<0$, both infinite in the $yz$-plane. As we consider spin waves with an in-plane wavevector $\mathbf{k}=(k_y,k_z)$, we analyze the system in 2D Fourier space. Because the fields applied in our work are much smaller than $\mu_0 M_s$, where $M_s$ is the YIG saturation magnetization, the equilibrium magnetization lies predominantly in-plane along $z$.

\subsection{The magnetic stray field of spin waves in a thin-film magnet}
The magnetic field generated by a magnetization $\mathbf{M}(\mathbf{r})=M_s \mathbf{m}(\mathbf{r})$, where $\mathbf{m}$ is a unit vector, is 
\begin{align}
\mathbf{B}(\mathbf{r}) &= \mu_0 M_s \int \mathbf{m}(\mathbf{r})\Gamma(\mathbf{r-r'})\mathrm{d}\mathbf{r}' 
\end{align}
where $\Gamma(\mathbf{r})$ is the dipolar tensor \cite{Kalinikos1986a}. Fourier transforming over the $y,z$ coordinates yields
\begin{align}
\mathbf{B}(\mathbf{k},x) &= \mu_0 M_s \int_{-t}^0 \Gamma(\mathbf{k},x-x')\mathbf{m}(\mathbf{k},x') dx' 
\label{eq:Bconv_z}
\end{align}
with $\Gamma(\mathbf{k},x)$ the dipolar tensor in 2D Fourier space, given by 
\be
\Gamma(\mathbf{k}, x) = - \frac{1}{2}ke^{-k|x|}  \left( \begin{array}{ccc} 
	2 \delta(x)/k  - 1 & i\sigma_x s  & i \sigma_xc\\
	i\sigma_x s  & s^2   & sc   \\
	i\sigma_x c   & s c   & c^2   \\
\end{array} \right) \label{eq:FFT_Gamma}
\ee
where we defined $\sigma_x =  \text{sign}(x)$, $s=k_y/k$, $c= k_z/k$, and $k=|\mathbf{k}|$. 

Spin waves are described by the dynamics of the transverse magnetization components $m_{x,y}$. Because the spin-wavelengths in our work are much larger than the YIG film thickness, we are in the 2D limit such that $\mathbf{m}(\mathbf{k},x) = \mathbf{m}(\mathbf{k})$, i.e., the magnetization dynamics are homogeneous over the film thickness. In this case, the out-of-plane component of the magnetic field above the film is 
\be
B_x(\mathbf{k},x) = \frac{\mu_0M_s}{2}(1-e^{-kt})  (m_x-i  s m_y) e^{-kx}, \label{eq:B0}
\ee
which we write as $ B_x^{sw}(\mathbf{k},x) =  B_0(\mathbf{k})e^{-kx}$. In the following, we omit the in-plane wave vector $\mathbf{k}$ from the argument lists for brevity. 

\subsection{The magnetic field generated by the spin-wave induced eddy currents}
The in-plane eddy currents in the metal film are induced by the out-of-plane component of the magnetic field according to Faraday's law: 
\be
-[\nabla \times \mathbf{E}]_x = -\frac{\partial E_z}{\partial y} +\frac{\partial E_y}{\partial z} = \frac{\partial B_x(\mathbf{r},t)}{\partial t},
\label{eq:FaradaysLaw}
\ee 
where $\mathbf{E}$ is the electric field. In Fourier space, this yields:
\begin{align}
	cJ_y(x)-	s J_{z}(x)  = \frac{\omega \sigma}{k} B_x(x) 
	\label{eq:eddycurrent1}
\end{align}
where $\mathbf{J} = \sigma\mathbf{E}$ with $\sigma$ the metal conductivity. 
We calculate the magnetic field generated by the current distribution $\mathbf{J}=(J_y,J_z)$ by expressing $\mathbf{J}$ as an effective out-of-plane magnetization \cite{Casola2018} using:
\be
M_s^\text{eff} m_x^\text{eff}(x) = -\frac{i}{k}(	cJ_y(x) - sJ_z(x))
\label{eq:m_eff}
\ee
The total field $B_x(x)$ is the sum of the eddy-current field and the spin-wave drive field
\be
B_x(x) = \mu_0M_s^\text{eff} \int_0^h \Gamma_{xx}(x-x') m_x^\text{eff}(x') dx' + B_x^{sw}(x)
\label{eq:BcNoAvg}
\ee
Substituting \eref{eddycurrent1} and the expression for $\Gamma_{xx}$ from \eref{FFT_Gamma} leads to the integral equation 
\begin{align}
B_x(x) &=\frac{-i\omega \mu_0 \sigma}{k^2}\int_0^h \frac{1}{2}ke^{-k|x-x'|}B_x(x')dx' + B_x^{sw}(x) 
\label{eq:BcNoAvg3}
\end{align}
which needs to be solved self-consistently. Note we excluded the delta function from $\Gamma_{xx}$ because we are analyzing free currents instead of spins.
 To solve \eref{BcNoAvg3}, we use a trial solution $B_x(x) = A_1B_0e^{-\kappa x} + A_2 B_0 e^{\kappa x}$. This gives 
 \begin{align}
 	A_1e^{-\kappa x} + A_2 e^{\kappa x} &=\frac{-i\omega \mu_0 \sigma}{2k^2}A_1\left(\frac{k}{k-\kappa}(e^{-\kappa x}-e^{-kx})+\frac{k}{k+\kappa}(e^{-\kappa x}-e^{-\kappa h}e^{k(x-h)})\right) \nonumber \\ &+ \frac{-i\omega \mu_0 \sigma}{2k^2}A_2\left(\frac{k}{k+\kappa}(e^{\kappa x}-e^{-kx})+\frac{k}{k-\kappa}(e^{\kappa x}-e^{\kappa h}e^{k(x-h)})\right) + e^{-kx}
 	\label{eq:BcNoAvg10}
 \end{align}
From equating the prefactors of the exponents with the same $x$-dependence, we get the solution 
\begin{align}
	\kappa & = \sqrt{ k^2+i\omega \mu_0 \sigma}, \nonumber \\
	A_1 & = A \frac{1}{1-\frac{a_-^2}{a_+^2}e^{-2\kappa h}}, \nonumber \\
	A_2 & = -A \frac{a_-}{a_+}e^{-2\kappa h} \frac{1}{1-\frac{a_-^2}{a_+^2}e^{-2\kappa h}}, 
	\label{eq:BcNoAvg11}
\end{align}
where we defined $A = \frac{2ik^2a_-}{\omega\mu_0\sigma}$ and $a_\pm = 1\pm\kappa/k$. 

Having solved for $B(x)$ inside the metal, we obtain the eddy currents from \eref{eddycurrent1}. As before, we calculate the field generated by the eddy currents by expressing these currents as an effective magnetization using \eref{m_eff}:
\be
\mu_0M_s^\text{eff}m_x^\text{eff}(x) = 2(1-\frac{\kappa}{k})\frac{A_1B_0e^{-\kappa x} + A_2 B_0 e^{\kappa x}}{A}
\ee
The eddy-current field above and below the metal is given by:
\begin{align}
B_x^e(x>h) &= \mu_0M_s^\text{eff} \int_0^h\frac{1}{2} k e^{-k(x-x')} m_x^\text{eff}(x')dx' =  \left(A_1(e^{a_-kh}-1) + A_2 \frac{a_-}{a_+} (e^{a_+kh}-1)\right) \frac{B_0}{A} e^{-kx} \\
B_x^e(x<0) &= \mu_0M_s^\text{eff} \int_0^h\frac{1}{2} k e^{k(x-x')} m_x^\text{eff}(x')dx' = \left(A_1\frac{a_-}{a_+}(1-e^{-2\kappa h}\right)\frac{B_0}{A} e^{kx} = \frac{-i\omega\mu_0\sigma}{k^2}\frac{1-e^{-2\kappa h}}{a_+^2-a_-^2e^{-2\kappa h}} B_0 e^{kx}
\label{eq:eddycurrentfield}
\end{align}
To calculate the coupling to the lowest-order perpendicular spin-wave mode studied in our work, we average \eref{eddycurrentfield} over $-t<x<0$, which corresponds to replacing $e^{kx}\rightarrow g_t = \frac{1-e^{-kt}}{kt}$. 

In the limit $2 |\kappa| h \ll 1$, \eref{eddycurrentfield} reduces to $B_x^e(x<0) = \frac{-i\omega\mu_0\sigma h}{2k} B_0e^{kx}$ as found in \cite{Yu2022}. However, in our experiments we are not in this limit as $\kappa>k$ because of the superconducting penetration depth (see next section). Furthermore, we observe that the expression reduces to $B_x^e(x<0) = -B_0e^{kx}$ when $|\omega\mu_0\sigma| \gg k^2$. In this limit the screening exactly cancels the drive field at $x=0$ and has become independent of $\sigma$.

\subsection{Decay of the magnetic field in a metal/superconductor thin film}
We calculate the decay constant $\kappa$ for a superconducting film. Using a two-fluid model, the conductivity of the superconductor is $\sigma = \sigma' - i\frac{1}{\omega L'}$, where $\sigma' = \sigma_n n_n(T)/n$, with $\sigma_n$ the normal-state conductivity of the metal, $n_n/n$ the fraction of non-cooper pair electrons, and $L'=\mu_0\lambda^2(T)$ is the specific kinetic inductance, with $\lambda_L$ the London penetration depth. Defining the skin depth $\delta = \sqrt{\frac{2}{\sigma_n\omega\mu_0}}$, we get
\be
\kappa = k\sqrt{1+i\omega\mu_0\sigma/k^2} = k\sqrt{1+\frac{2i}{k^2\delta^2(T)}\frac{n_n(T)}{n}+\frac{1}{k^2\lambda_L^2(T)}}
\label{eq:kappa}
\ee
The imaginary term underneath the square root induces additional spin-wave damping \cite{Bertelli2021}, but this effect is small because the MoRe skin depth $\delta \approx 4$$ \mu$m for $\omega/2\pi = 3 $ GHz far exceeds the MoRe film thickness and because the fraction of normal electrons $n_n(T)$ decreases quickly with decreasing temperature \cite{Tinkham2004}. As such, we neglect the real part of the conductivity when calculating the hybidized spin-wave dispersion below. 

\subsection{Comparison geometric and kinetic inductance}
To assess the role of geometric inductance in our measurements, we first consider the energy cost of creating a sinusoidal current pattern in the superconducting strip $J_z(y) = J_0\cos(k_0y)\cos(\omega t)$. The kinetic energy density of the supercurrent is $\mu_0\lambda_L^2J_z^2/2$ \cite{Clem2013}. The kinetic inductance per unit length is defined by 
\be
\frac{1}{2} L_k I_z^2=\frac{1}{2}\mu_0\lambda_L^2\int J_z^2dxdy
\ee
and depends on temperature through $\lambda_L$. The geometric inductance per unit length is defined by 
\be
\frac{1}{2}L_gI_z^2 = \frac{1}{2\mu_0}\int B^2 dx dy
\ee
with $I_z = I_0\cos(\omega t)$ the current running in the $z$ direction, where $I_0=\tfrac{2WJ_0h}{\pi}$. The geometric inductance is independent of temperature. Neglecting the finite thickness of the superconductor, the magnetic field generated by $J_z$ is
\begin{align}
	B_x &= -\frac{\mu_0}{2}J_0h e^{-k_0|x|}\sin(k_0y)\cos(\omega t) \nonumber \\
	B_y &= \frac{\mu_0}{2}J_0h e^{-k_0|x|}\cos(k_0y)\cos(\omega t).
\end{align}
It follows that the ratio of the kinetic and geometric inductance is
\be
\frac{L_k}{L_g} = \mu_0^2\lambda_L^2\frac{\int J_z^2dxdy}{\int B^2 dxdy} = k_0\Lambda 
\ee
where we assumed $k_0W\gg 1$ so that we can neglect edge effects and where $\Lambda = 2\lambda_L^2/h$ is the Pearl length \cite{Brandt1993}. Considering $\lambda_L > 0.4$ $\mu$m and spin-wavenumbers $k_0 > 0.8 $ $\mu$m$^{-1}$ (c.f. Fig.~4b and black line in Fig.~4c of the main text), we get $ k_0\Lambda > 2$. We observe that the kinetic inductance dominates. 

To assess the inductance for the eddy-current pattern excited by the spin waves, we now consider the traveling-wave eddy-current pattern $J_z(y,t) = J_0\cos(ky - \omega t)$ running in an infinite metal film. In this case, the total current running in the $+z$ direction is independent of time, as is the energy density of the associated magnetic field:
\be
\frac{1}{2\mu_0} (B_x^2+B_y^2) = \frac{\mu_0}{2}  \frac{(J_0h)^2}{4} e^{-2k_0|x|}
\ee
This situation resembles the time-independent magnetic field generated by a DC current in a wire: the geometric inductance does not impede the motion of the eddy current pattern. However, at the edges of the superconducting strip, this does not hold and both the current and the magnetic-field energy density vary in time. Considering the effect of the edges only, the time-dependent part of the total current is, at most, $I(t) = I_1\cos(\omega t)$, with $I_1=J_0h \frac{2}{k_0}$. Considering the $W=30$ $\mu$m width of the superconducting strip, we thus expect the role of $L_g$ to be suppressed by a factor $I_1/I_0= k_0W/\pi \gg 1$, and the kinetic inductance to dominate by a factor $ k_0^2\Lambda W \gg 1 $. 

\subsection{Dispersion of spin waves interacting with a superconductor}
In this section we derive the dispersion of spin waves underneath a superconducting film. We do so by including the field generated by the spin-wave induced Meissner currents self-consistently into the LLG equation.

We start by deriving the spin-wave dispersion in the absence of the superconductor. The LLG equation is
\be
\dot{\mathbf{m}} = -\gamma \mathbf{m}\times \mathbf{B} - \alpha \dot{\mathbf{m}}\times \mathbf{m}
\label{eq:LLGeq}
\ee	
where $\gamma$ is the electron gyromagnetic ratio, $\mathbf{B}$ is the effective magnetic field, and $\alpha$ the Gilbert damping parameter. Linearized and in the frequency domain, it becomes
\begin{align}
i\omega m_{x} = \gamma(m_{y} B_{z} - B_{y})  -i\alpha \omega m_{y} \nonumber \\
i\omega  m_{y} = \gamma(B_{x}-m_{x} B_{z} ) + i\alpha \omega m_{x}  
\label{eq:LLGlin2}
\end{align}
where the linearization implies $m_{z}=1$ and that $B_{z}$ is the static part of the $z$-component of the magnetic field while both $B_{x,y}$ oscillate at frequency $\omega$.
By calculating the effective magnetic field associated with the applied field, the exchange coupling, and the demagnetizing field \cite{Rustagi2020}, we get 
\begin{align}
\gamma B_{x} &  =\omega_{M}(f-1)m_{x}-Dk^{2}m_{x}+\gamma
B_{\mathrm{AC},x},\nonumber \\
\gamma B_{y} &  =-\omega_{M}fs^{2} m_{y}-Dk^{2}m_{y}+\gamma
B_{\mathrm{AC},y}
\label{eq:Bz_sol}
\end{align}
where we defined $f =  1 - g_t$, $D$ is the spin stiffness, and $\omega_M =  \gamma \mu_0 M_s$. Solving for the spin-wave dispersion $\omega_{sw}$ of the bare-YIG film, we find 
\be
\omega_{sw} =\sqrt{\omega_2\omega_3}, 
\label{eq:bareYIGdispersion}
\ee 
with
\begin{align}
\omega_{0} &  =\omega_{B}+\omega_{D}k^{2} , \nonumber\\
\omega_{2} &  =\omega_{0}+\omega_{M}(1-f) , \nonumber \\
\omega_{3} &  =\omega_{0}+\omega_{M}fs^{2} ,\label{eq:omega3}
\end{align}
where we defined $ \omega_B = \gamma B_z$ with $B_z = \cos(\theta)B_{NV}$ the in-plane component of the applied magnetic field.

\subsubsection{Meissner current contribution to the LLG-equation}
To calculate the spin-wave dispersion in our YIG-superconductor hybrid, we incorporate the magnetic field generated by the spin-wave induced eddy currents into the LLG equation. The spin waves generate a magnetic field $B_x=B_0e^{-kx}$, with (see \eref{B0})
\be
B_0 = \frac{\mu_0 M_s k t}{2}g_t(m_x-ism_y)
\ee
Substituting into \eref{eddycurrentfield} yields the eddy-current-generated field in the magnetic film. Averaging this field over $-t<x<0$ to calculate the coupling to the lowest-order perpendicular spin-wave mode leads to
\begin{align}
\gamma \overline{B}_{e,x} = & i \omega \alpha_{m} (m_{x} -i s m_{y}), \nonumber \\
\gamma \overline{B}_{e,y} = & -\alpha_{m}\omega s (m_{x} -i s m_{y})
\label{eq:eddycurrentbackactionfield},
\end{align}
where 
\be
\alpha_{m} =  -\frac{\gamma \mu_0^2 M_s \sigma k t}{2k^2} g_t^2 \frac{1-e^{-2\kappa h}}{a_+^2-a_-^2e^{-2\kappa h}}
\label{eq:eddycurrentdamping}
\ee
is a dimensionless factor that alters the spin-wave dispersion as discussed in the next section. Assuming the kinetic inductance to dominate (see section above), we have $\sigma = \frac{-i}{\omega \mu_0 \lambda_L^2}$ so that
\be
\alpha_{m} =  i\frac{\omega_L}{\omega} \quad \text{with} \quad \omega_L = \frac{\gamma \mu_0 M_s  k t}{2k^2\lambda_L^2} g_t^2 \frac{1-e^{-2\kappa h}}{a_+^2-a_-^2e^{-2\kappa h}}
\label{eq:wL}
\ee
We note that $\omega_L\rightarrow 0$ for both $k\rightarrow 0 $ and $k\rightarrow \infty $ as the spin-wave stray field vanishes in this limit.

\subsubsection{Spin-wave dispersion underneath the superconductor}
To derive the spin-wave dispersion underneath the superconductor, we add the eddy-current field of \eref{eddycurrentbackactionfield} to the effective magnetic field  
\begin{align}
\gamma B_{x} &  =-(\omega_{M}(1-f)+\omega_{D}k^{2})m_{x}+i\omega\alpha_m\left(
m_{x}-ism_{y}\right)  +\gamma B_{AC,x},\nonumber \\
\gamma B_{y} &  =-(\omega_{M}f+\omega_{D}k^{2})m_{y}-\alpha_m\omega s\left(
m_{x}-i s m_{y}\right)  +\gamma B_{AC,y},
\label{eq:field_inplane}%
\end{align}
such that the linearized LLG equation \eref{LLGeq} becomes  
\begin{align}
-i\omega m_{x} &  =-(\omega_{3}-i(s^2\alpha_m+\alpha)\omega)m_{y}-\alpha_m s \omega m_{x}+\gamma
B_{AC,y}, \nonumber \\
-i\omega m_{y} &  =(\omega_{2}-i(\alpha_m+\alpha)\omega)m_{x}-\alpha_m\omega s m_{y}-\gamma
B_{AC,x},
\end{align}
where $\omega_{2}$ and $\omega_{3}$ are given in \eref{omega3}). In matrix form:
\begin{equation}%
\begin{pmatrix}
\omega_{2}-i(\alpha+\alpha_m)\omega & (i-\alpha_m s) \omega \\
-(i-\alpha_m s)\omega & \omega_{3}-i(\alpha+\alpha_m s^2)\omega
\end{pmatrix}%
\begin{pmatrix}
m_{x}\\
m_{y}%
\end{pmatrix}
=\gamma%
\begin{pmatrix}
B_{AC,x}\\
B_{AC,y}%
\end{pmatrix}.
\label{eq:LLG_matrix_eddy}%
\end{equation}
The resulting susceptibility is singular when 
\be
(\omega_{2}-i(\alpha+\alpha_m)\omega)(\omega_{3}-i(\alpha+s^2\alpha_m)\omega) + (i-s\alpha_m)^2\omega^2=0.
\ee
Splitting $\alpha_m$ into its real and imaginary parts via $\alpha_m=\beta+i\zeta$ and disregarding 2nd-order terms in $\alpha$ and $\beta$, we get
\begin{align}
\omega_2\omega_3+\omega\zeta(\omega_2s^2+\omega_3)-\omega^2(1-2s\zeta) = 0
\label{eq:realLambda}
\end{align}
We recall that  $\alpha_m \approx i\zeta=i\omega_L/\omega$, with $\omega_L$ given by \eref{wL}. Substituting into \eref{realLambda}, we get
\begin{align}
	\omega_2\omega_3+\omega_L(\omega_2s^2+\omega_3)-\omega^2+2s\omega_L\omega = 0
\end{align}
where we disregarded 2nd-order terms in $\alpha$. Solving this equation leads to 
\be
\omega_{sw}=s\omega_L\pm \sqrt{\omega_2\omega_3}\sqrt{1+\frac{\eta^2s^2+1}{\eta}\frac{\omega_L}{\sqrt{\omega_2\omega_3}}+\frac{s^2\omega_L^2}{\omega_2\omega_3}}
\ee
where $\eta=\sqrt{\omega_3/\omega_2}$. Neglecting terms of order $\omega_L^2/\omega_2\omega_3$ and disregarding the negative-frequency solution leads to 
\be
\omega_{sw}=\sqrt{\omega_2\omega_3} + \omega_L\frac{(\eta s+1)^2}{2\eta}
\label{eq:SC_dispersion}
\ee
which yields Eq. 1 of the main text $f_\text{YIG/SC} = f_\text{YIG} + f_\text{SC}$, where
$f_\text{YIG} = \sqrt{\omega_2\omega_3}/2\pi$ is the bare-YIG dispersion and
\be
f_\text{SC} = \frac{\gamma \mu_0 M_s  k t}{2k^2\lambda_L^2} g_t^2 \frac{1-e^{-2\kappa h}}{a_+^2-a_-^2e^{-2\kappa h}} \frac{(\eta s+1)^2}{2\eta} \approx \gamma \mu_0 M_s kt r \frac{1-e^{-2h/\lambda_L }}{(k\lambda_L+1)^2-(k\lambda_L-1)^2 e^{-2h/\lambda_L } }
\ee
is the superconductor-induced shift, where we defined the dimensionless geometrical factor $r=\tfrac{g_t^2 }{2}\frac{(\eta s+1)^2}{2\eta}$, and where the approximation holds when the kinetic inductance dominates the impedance and when $k^2\lambda_L^2 \ll 1$ (c.f. \eref{kappa} and Figure 4B-C of the main text).

\section{Extracting $M_s$ and $\lambda_L$.}
Here we describe the extraction of the saturation magnetization $M_s$ of the YIG film and the London penetration depth $\lambda_L$ of the MoRe strip. We determine $M_s$ from the magnetic-field dependence of the spin-wavenumber $k_\text{YIG}$ in the bare-YIG region ($y<0$ region of Fig.~3a-b of the main text). $M_s$ does not significantly change over the $T$ = 5.5-10 K temperature range in our measurements such that the spin-wave length is temperature-independent in the bare-YIG region (Fig.~\ref{fig:S5}). Knowing $M_s$, we subsequently determine $\lambda_L$ from the magnetic-field dependence of the spin-wavenumber $k_\text{YIG/SC}$ in the YIG+MoRe region ($y>0$ region of Fig.~3a-b of the main text). 

To extract the wavenumbers  $k_\text{YIG}$ and $k_\text{YIG/SC}$, we separate our spatial maps of the NV ESR contrast $C$ into a bare-YIG and a YIG+MoRe region, and then linearly detrend the data along the $y$-direction (Fig.~S2a-b). We pad the data with zeros, Fourier transform along $y$ (Fig.~S2d-e), and extract $k_\text{YIG}$ for the bare-YIG and $k_\text{YIG/SC}$ for the YIG+MoRe region by finding the FFT amplitude maximum (Fig.~S2f). The uncertainty $\epsilon^{(h)}$ of the extracted wavenumbers is determined by the inverse of the spatial sampling range $1/\Delta y^{(h)}$. The sampling ranges are indicated in Fig.~S2c.

From the field dependence of the spin-wave number in the bare-YIG region, we extract the saturation magnetization $M_s$ via least-squares minimization of 
\be
\chi^2_j = \sum_{i=1}^N(f_{sw}(B_z(i), k_\text{YIG}^{(j)}(i), M_s) - f_{NV_{j}}(i))^2.
\ee
Here, $f_{sw}=\omega_{sw}/2\pi$ is the spin-wave dispersion in the bare-YIG region given by \eref{bareYIGdispersion}, $i=1,2...N$ indexes the measurements at the different magnetic fields $B_z(i)$, $k_\text{YIG}^{(j)}(i)$ are the corresponding extracted wavenumbers (see previous subsection), $M_s$ is a free parameter, and $j = 1$ ($j = 2$) indicates the field-aligned (misaligned) NV ensemble used in Fig.~3a (Fig.~3b) of the main text.

Having found $M_s$, we extract $\lambda_L$ using a similar procedure, by least-squares minimization of
\be
\chi^2_j = \sum_{i=1}^N(f_{sw}(B_z(i), k_\text{YIG/SC}^{(j)}(i), M_s, \lambda_L) - f_{NV_{j}}(i))^2.
\ee
Here, $f_{sw}=\omega_{sw}/2\pi$ is the spin-wave dispersion in the YIG/MoRe region given by \eref{SC_dispersion}, $M_s$ is fixed, and $\lambda_L$ is a free parameter. Additional parameters we keep fixed in the fitting procedures are the thicknesses of the YIG and MoRe films.

\begin{figure*}[!ht]
\includegraphics{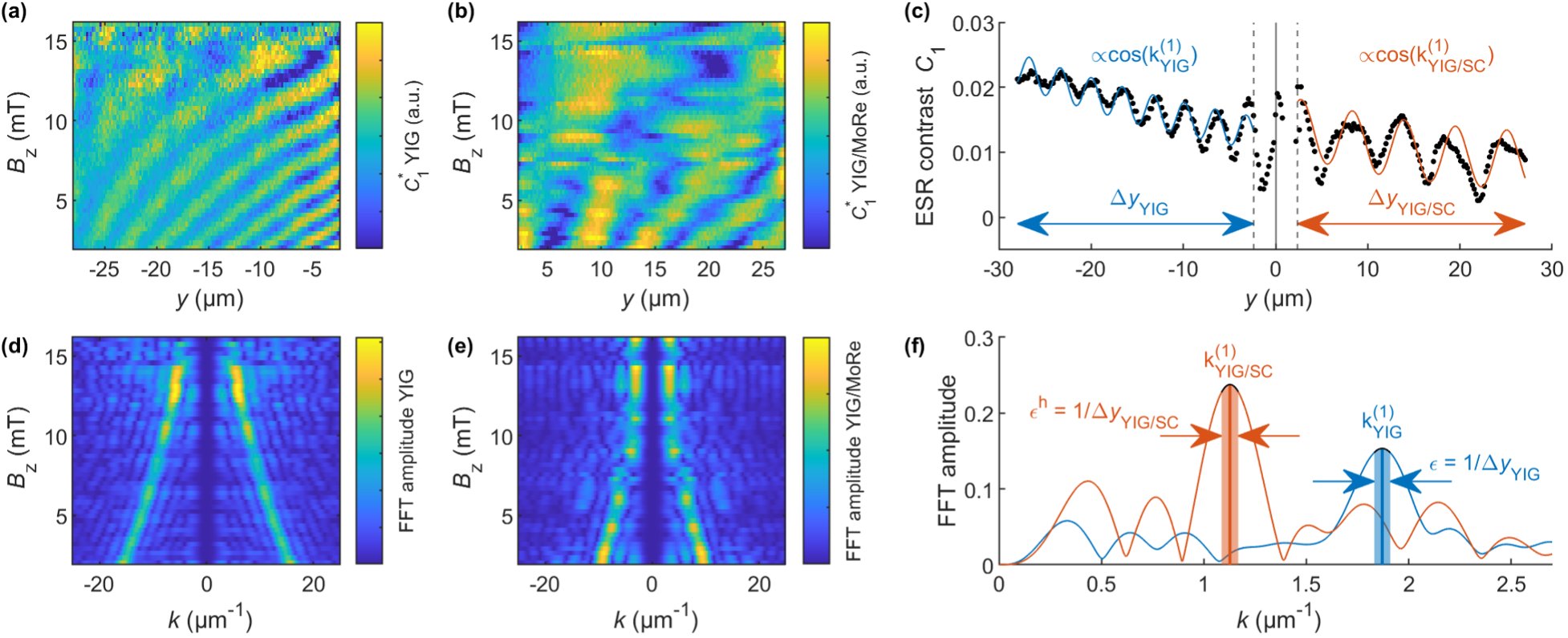}
\caption{\label{fig:S2} \textbf{Extracting the spin-wavenumber through Fourier analysis.} (a-b) Detrended and line-by-line normalized field-dependent ESR contrast for the YIG and YIG/MoRe regions. (c) ESR contrast as a function of position, example of raw data. The solid vertical line indicates the border between bare magnet ($y<0$) and the heterostructure ($y>0$), we ignore data between the dashed lines in our fitting process. (d-e) Fourier transform of data in (a-b) padded with zeros. (f) Fourier transform of data in (c). The wavenumbers $k_\text{YIG}$ and $k_\text{YIG/SC}$ are determined by finding the peak location in the Fourier spectrum. The uncertainty $\epsilon(h)$ (indicated by the shading) is determined by the inverse of the sampling range $\Delta y_{YIG(/SC)}$. }
\end{figure*}

\begin{figure*}[!ht]
\includegraphics{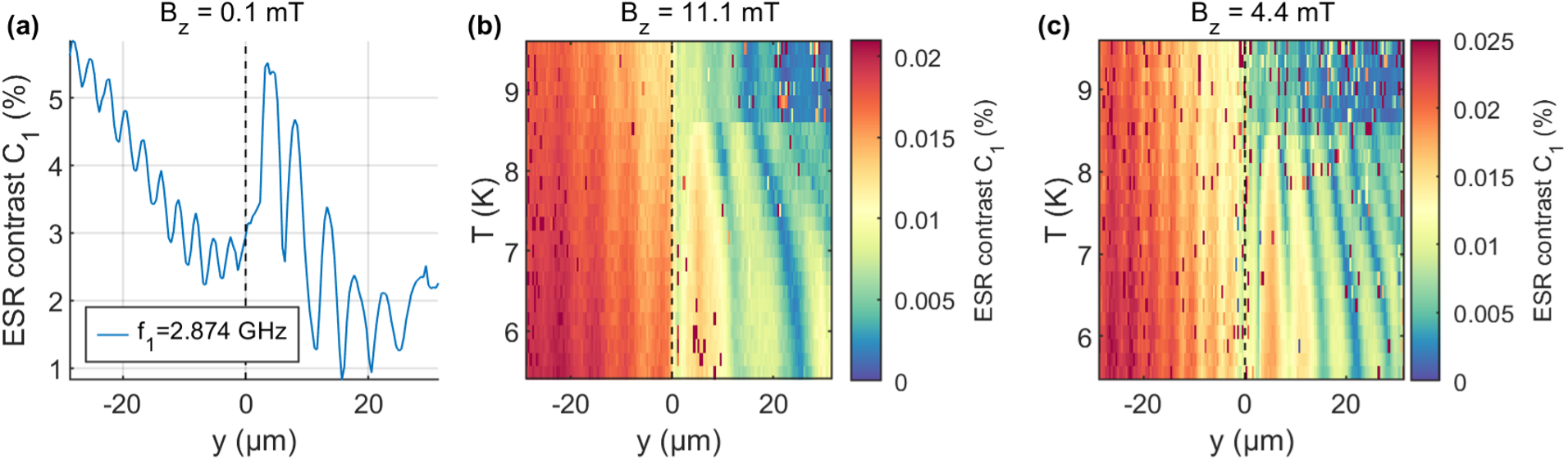}
\caption{\label{fig:S3} \textbf{Observation of the wavelength shifts with purely in-plane magnetic fields.} (a) ESR contrast as a function of position at $B_z$ = 0.1 mT and $T$ = 5.5 K, showing hybrid modes with stretched wavelength for $y > 0$. Measurements are taken with a field applied in-plane ($\theta=0$ deg., such that $B_z=B_\text{NV}$). (b, c) Spin waves in YIG and YIG/MoRe for two different, purely in-plane fields $B_z = 11.1$ mT (b) and $B_z = 4.4$ mT (c), with stretched-wavelength modes for $y > 0$ and $T < T_c$. As the applied field is in-plane, Meissner screening of this field is negligible (i.e., it does not introduce a static field inhomogeneity that could affect the spin-wave dispersion). }
\end{figure*}

\begin{figure*}[!ht]
\includegraphics{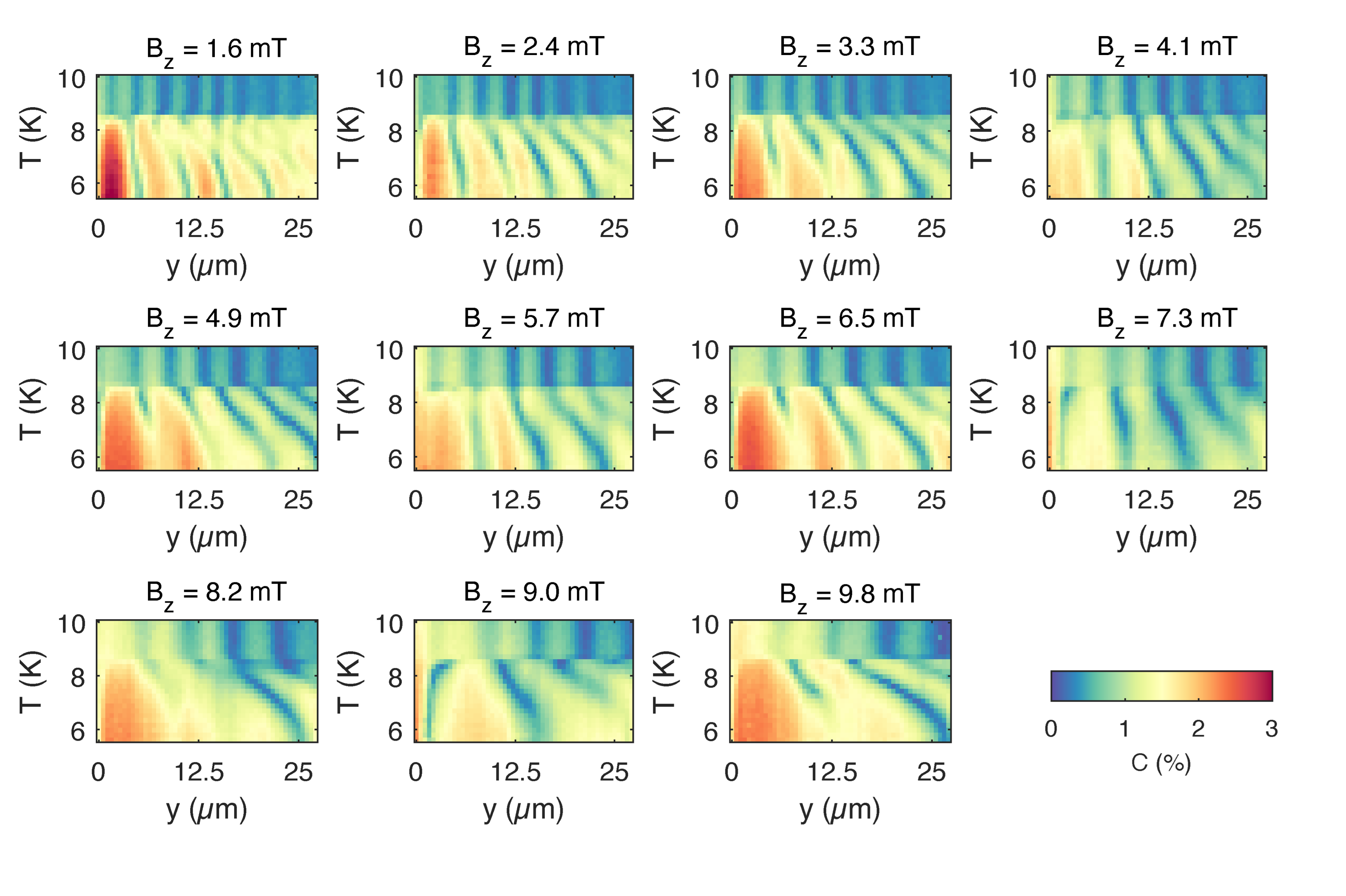}
\caption{\label{fig:S4} \textbf{Temperature dependence of the wavelength shift for various magnetic fields.} Linetraces of the NV ESR contrast in the YIG/MoRe region as a function of temperature. Using Fourier analysis, we extract the wavenumbers as a function of temperature shown in Fig.~4b-c of the main text. As we cool down through the transition temperature, we observe a continuous decrease in the wavenumber in all measurements. Below Tc we observe a global increase in the ESR contrast, presumably caused by an increase in inductively coupled current in the MoRe strip.}
\end{figure*}

\begin{figure*}[!ht]
\includegraphics{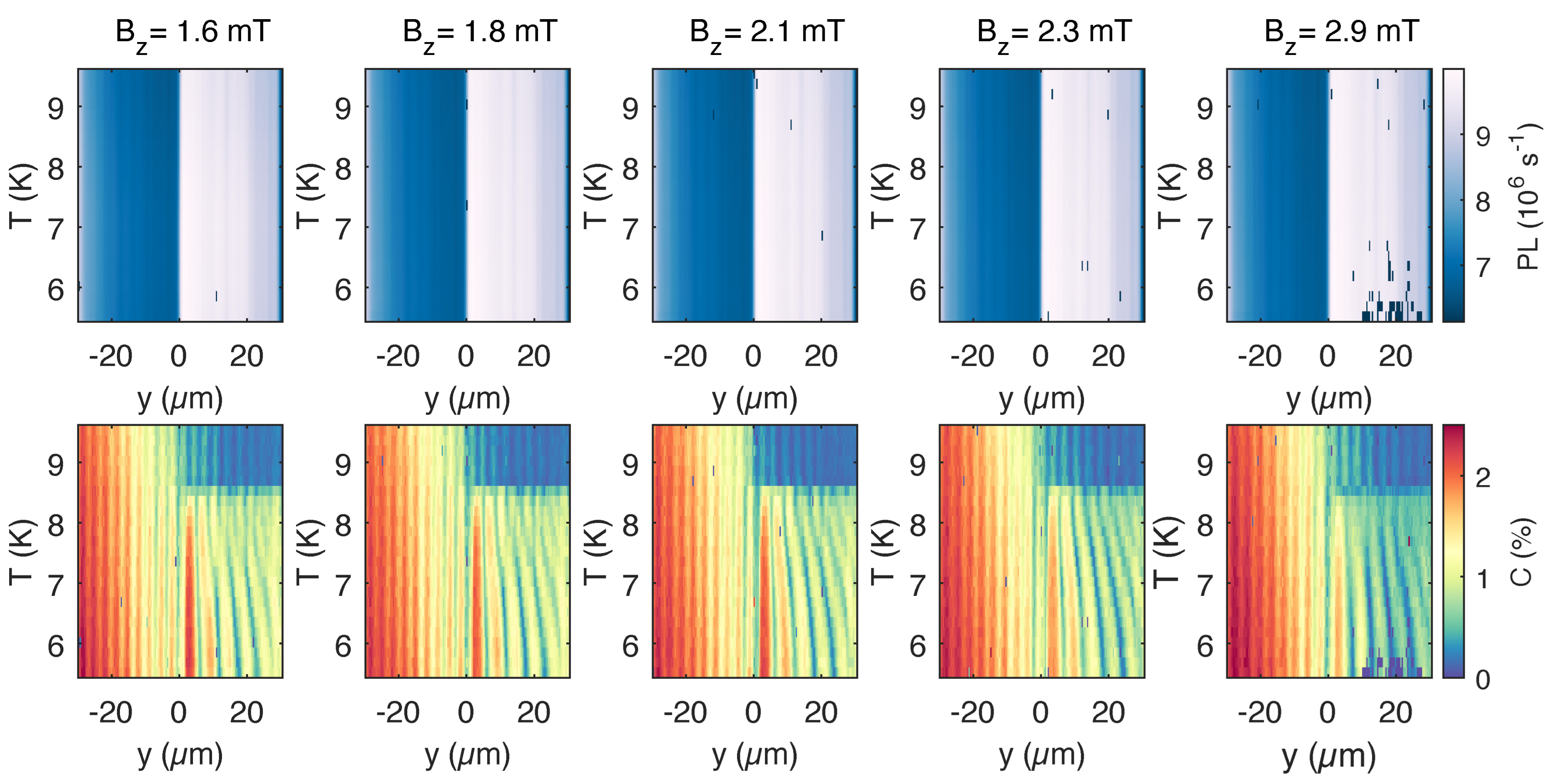}
\caption{\label{fig:S5} \textbf{Spin-wave length as a function of temperature for various magnetic fields $B_z$}. Top row of panels: NV photoluminescence (PL) showing the location of the MoRe strip at y > 0. Bottom row of panels: Linetraces of the NV ESR contrast across the MoRe strip as a function of temperature, showing the temperature-dependent shift in wavenumber in the YIG/MoRe region ($y > 0$, $T < T_c$ ) while the wavenumber in the bare-YIG region ($y < 0$) remains unchanged..}
\end{figure*}

\begin{figure*}[!ht]
\includegraphics{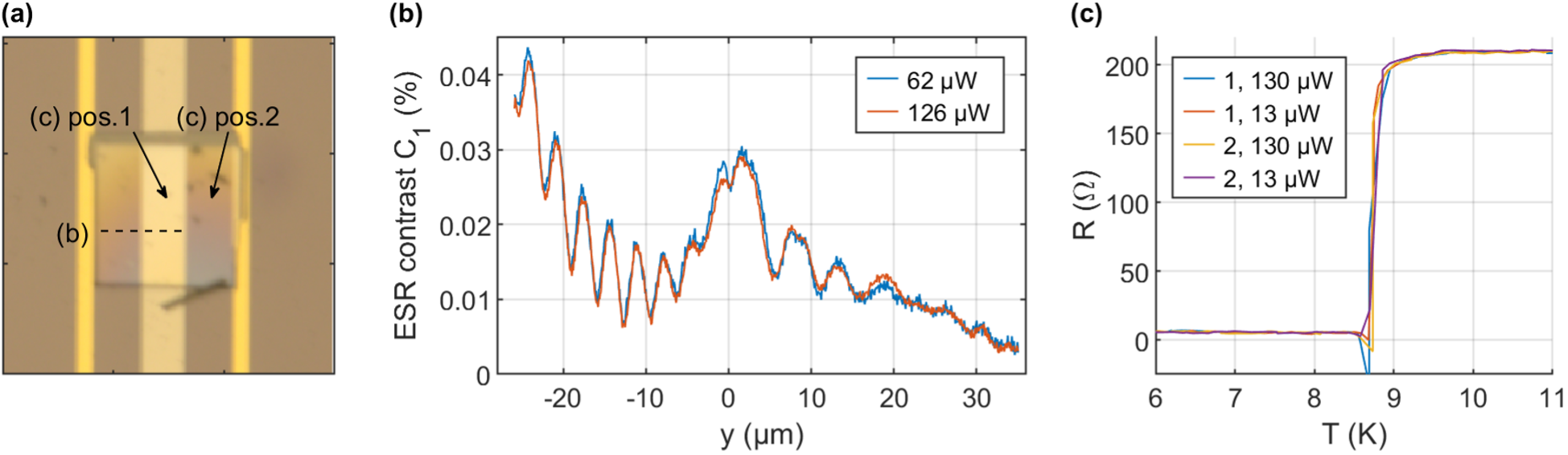}
\caption{\label{fig:S6} \textbf{Investigation of global heating of the superconductor due to NV excitation laser}. (a) To exclude the possible influence of laser heating on the extracted spin-wave lengths, we image the spin wavelength and measure the DC resistance of the MoRe strip as different laser powers and laser positions. The dashed line in the microscope image indicates the location of the ESR linetraces in panel (b). The arrows indicate the laser location in the resistance measurements of panel (c). (b) Spatial line trace of the NV ESR contrast at $T$ = 5.5 K with the power of our 520 nm NV excitation laser set at 62 µW and 126 µW. We do not observe a significant difference in spin-wave lengths (c) DC transport resistance of the superconducting strip at different laser powers, with the laser focused at the two locations indicated in (a). We do not observe a significant difference in $R|_{T<T_c}$ and $T_c$ between the measurements. }
\end{figure*}

\clearpage
\bibliography{export.bib}

%% file: sw_sc_abstract.tex
Superconductors are materials with zero electrical resistivity and the ability to expel magnetic fields known as the Meissner effect. Their dissipationless diamagnetic response is central to magnetic levitation and circuits such as quantum interference devices. Here, we use superconducting diamagnetism to shape the magnetic environment governing the transport of spin waves – collective spin excitations in magnets that are promising on-chip signal carriers – in a thin-film magnet. Using diamond-based magnetic imaging, we observe hybridized spin-wave–Meissner-current transport modes with strongly altered, temperature-tunable wavelengths. We extract the temperature-dependent London penetration depth from the wavelength shifts and realize local control of spin-wave refraction using a focused laser. Our results demonstrate the versatility of superconductor-manipulated spin-wave transport and have potential applications in spin-wave gratings, filters, crystals and cavities.